\documentclass[fleqn,usenatbib]{mnras}

\usepackage{newtxtext,newtxmath}

\usepackage[T1]{fontenc}

\DeclareRobustCommand{\VAN}[3]{#2}
\let\VANthebibliography\thebibliography
\def\thebibliography{\DeclareRobustCommand{\VAN}[3]{##3}\VANthebibliography}

\usepackage{graphicx}	
\usepackage{amsmath}	

\usepackage{color}
\usepackage[normalem]{ulem}

\title[Alfvén wave propagation, reflection and trapping in the solar wind]{Alfvén wave propagation, reflection and trapping in the solar wind}

\author[A. Kumar et al.]
{
A. Kumar,$^{1}$\thanks{E-mail: ak394@st-andrews.ac.uk}
T. A. Howson,$^{2, 1}$
P. Pagano$^{3, 4, 1}$
and I. De Moortel$^{1, 5}$
\\
$^{1}$School of Mathematics and Statistics, University of St Andrews, St Andrews, Fife KY16 9SS, UK\\
$^{2}$Faculty of Design, Informatics and Business
Abertay University, Bell Street, Dundee, DD1 1HG, UK
\\
$^{3}$Dipartimento di Fisica \& Chimica, Università degli Studi di Palermo, Via Archirafi 36, I-90123 Palermo, Italy\\
$^{4}$INAF-Osservatorio Astronomico di Palermo, Piazza del Parlamento 1, I-90134 Palermo, Italy\\
$^{5}$Rosseland Centre for Solar Physics, University of Oslo, PO Box 1029 Blindern, NO-0315 Oslo, Norway
}

\date{Accepted XXX. Received YYY; in original form ZZZ}

\pubyear{\the\year{}}

\begin{document}
\label{firstpage}
\pagerange{\pageref{firstpage}--\pageref{lastpage}}
\maketitle

\begin{abstract}
Alfvén waves are known to be important carriers of magnetic energy that could play a role in coronal heating and/or solar wind acceleration. As these waves are efficient energy carriers, how they are dissipated still remains one of the key challenges. Using a series of 1.5-D magnetohydrodynamic (MHD) simulations, we explore wave energy trapping associated with field-aligned density enhancements. We examine the parameters which govern the wave reflection and trapping. The goal of our simulations is to find optimal conditions for wave trapping, which would ultimately promote the energisation of the solar atmosphere. In agreement with previous studies, we find that maximum wave reflections happen only for a narrow range of density enhancement widths, namely when it is comparable to the Alfvén wave wavelength. In our paper, we explain this scale-selectivity using a semi-analytical model that demonstrates the importance of wave interference effects. As expected, we find that spatially extended regions of density inhomogeneities favour enhanced wave reflection and trapping. However, wave interference causes saturation of the reflected energy for very extended regions of varying density. 

\end{abstract}

\begin{keywords}
MHD -- waves -- Sun: solar wind -- Sun: oscillations
\end{keywords}

\section{Introduction}
Understanding the physical processes behind coronal heating and solar wind acceleration remains one of the key challenges in solar physics. As magnetohydrodynamic (MHD) waves are ubiquitously observed from different space-based missions in different parts of the solar atmosphere \cite[e.g.][and references therein]{cirtain2007evidence, de2007chromospheric, banerjee2009signatures, jess2015multiwavelength, bale2019highly, banerjee2021magnetohydrodynamic}, they are considered to be important and efficient carriers of magnetic energy from the solar interior to its atmosphere. Of the three fundamental MHD wave modes -- slow, fast and Alfvén -- Alfvén waves are considered to be the most efficient carriers because of their incompressible nature, due to which they can propagate up to few tens of solar radii as observed in-situ either as propagating Alfvénic fluctuations \citep[][]{belcher1971large} or in the form of magnetic switchbacks by the Parker Solar Probe (PSP) \citep[e.g.][]{kasper2019alfvenic, horbury2020sharp}. As a result, the role of Alfvén waves in coronal heating and solar wind acceleration has garnered significant scientific interest \citep[e.g.][]{cranmer2007self, mcintosh2011alfvenic, van2016heating}.

For Alfvén waves to contribute to the energisation of the solar atmosphere, the carried magnetic energy must be converted to thermal energy at appropriate locations. Depending on the large-scale magnetic field under consideration -- such as coronal loops, null points, coronal holes, etc. -- different theories of wave energy dissipation have been proposed, most of which are aimed at generating small length scales (typically of the order of a few metres, see for example \cite{judge2024problem}). It is only at such small scales that resistive and viscous effects dominate and wave energy dissipation becomes significant. Some of these theories include resonant absorption \citep[][]{ionson1978resonant}, phase mixing \citep[][]{heyvaerts1983coronal} and the Kelvin-Helmholtz instability in oscillating flux tubes (e.g. \citet{terradas2008nonlinear} or reviewed in \cite{howson2022khi}). These processes typically depend on perpendicular (to the field) inhomogeneities in the local Alfvén speed \citep[e.g., see][for a review]{morton2023alfvenic}. Additionally, the possibility of wave energy dissipation via turbulent cascade as a consequence of non-linear self-interaction of partially-reflected Alfvén waves has been explored \citep[e.g.][and references therein]{cranmer2005generation, verdini2007alfven, chandran2009alfven, van2011heating, perez2013direct, van2014alfven}. 

Coronal holes are known to be the source of fast solar wind, which can accelerate up to $\sim 500$-$700\, \mathrm{km\,s^{-1}}$ at $1\, \mathrm{AU}$. Because of the open magnetic field configuration in coronal holes, ions and electrons can propagate rather efficiently into space. This results in a relatively uniform and less variable structure of the fast solar wind \citep[e.g., see][for a review]{cranmer2009coronal}. Given the lack of (significant) cross-field gradients in such a scenario, dissipation by Alfvén wave turbulence becomes more important, which is commonly observed in the interplanetary medium \citep[][]{bale2005measurement, bandyopadhyay2020enhanced}.

A key consideration for a turbulent cascade of Alfvén wave energy is the reflection and trapping of wave energy at appropriate locations. Wave reflections usually occur in the presence of strong density gradients, which are readily available for coronal loops anchored in the chromosphere. This leads to wave trapping and the formation of standing modes, thereby increasing the opportunities for wave energy dissipation. On the other hand, along coronal holes (regions of open magnetic field), Alfvén waves can only undergo partial reflections due to the gravitationally stratified medium, unless there are strong density variations in the direction of the guide field. As counterpropagating Alfvén waves have been observed in coronal holes in Coronal Multi-Channel Polarimeter (CoMP) observations \citep[][]{morton2015investigating}, the possibility of Alfvén wave reflection and trapping due to field-aligned density variations is considered in recent studies \citep[e.g.][]{asgari2021effects, pascoe2022propagating}.

Non-linear Alfvén waves can produce density fluctuations as they propagate, either via ponderomotive forces \citep[][]{verwichte1999evolution}, or due to spherical expansion in the solar atmosphere \citep[][]{nakariakov2000nonlinear}. Such density fluctuations provide additional sites for wave reflection and trapping, possibly facilitating a turbulent cascade of wave energy into smaller length scales. Apart from this, (linear) Alfvén waves can also reflect from pre-existing density inhomogeneities in the solar atmosphere such as spicules and jets, resulting from the dynamic nature of the solar atmosphere.  

In the work presented in this paper, we study Alfvén wave reflection and trapping along open field structures, such as coronal holes, using 1.5-D MHD simulations. As in \cite{pascoe2022propagating}, we impose a density enhancement on top of a uniform medium. In \cite{pascoe2022propagating} and \cite{yuan2015evolution}, the authors have shown that reflections maximise when the density inhomogeneity length scale is roughly half the injected wave wavelength. This holds true for both Alfvén waves, as well as for fast magnetoacoustic waves. In this paper, we not only present cases for wave reflection and trapping with a background plasma flow, we also present a semi-analytical model that explains the scale selectivity seen in such studies. The objective of our study is to establish favourable conditions for Alfvén wave energy trapping in between density enhancements -- which are either uniformly spaced and identical, or randomly generated -- in terms of the different length scales in the system, the density contrast, and the background wind speed. Although we do not study wave heating directly, we do provide estimates on the energy budgets available for dissipation and heating. 

The paper is organised as follows. In Section \ref{section_numerical_setup}, we describe the model and the parameter space; results of which are described in Section \ref{section_results}. Furthermore, we construct a semi-analytical model that explains the scale-selectivity seen in our simulations, which is described in Section \ref{section_semi-analytical_model}. Lastly, we provide discussion and conclusions in Section \ref{section_discussion}.

\section{Numerical setup}\label{section_numerical_setup}

The simulations presented in this article were implemented using PLUTO -- a finite volume, shock-capturing code for solving the following equations in ideal MHD \citep{mignone2007pluto}. 
\begin{gather}
    \frac{\partial \rho}{\partial t} + \mathbf{\nabla}\cdot\left( \rho \mathbf{v}\right) = 0, \label{continuity}\\
    \rho \frac{\partial \mathbf{v}}{\partial t} + \rho \left( \mathbf{v}\cdot \mathbf{\nabla}\right) \mathbf{v} = -\mathbf{\nabla}p + \left( \mathbf{\nabla} \times \mathbf{B} \right)\times \mathbf{B}, \label{momentum}\\
    \frac{\partial \mathbf{B}}{\partial t} = \mathbf{\nabla} \times \left( \mathbf{v} \times \mathbf{B}\right), \label{induction}\\
    \frac{\partial p}{\partial t} + \mathbf{v}\cdot \mathbf{\nabla}p = -\gamma p \mathbf{\nabla}\cdot \mathbf{v}, \label{energy}\\
    p = nk_\mathrm{B}T. \label{eqn_of_state}
\end{gather}
Here, $\rho$, $p$, $\mathbf{v}$ and $\mathbf{B}$ denote the density, gas pressure, velocity and magnetic field, respectively. The ideal gas law (\ref{eqn_of_state}) provides closure to the system of equations, where $n$ denotes the number density, $k_\mathrm{B}$ is the Boltzmann constant and $T$ is the temperature. 

The above equations are advanced on a uniformly spaced one-dimensional grid aligned with the $x-$axis in Cartesian geometry using a third-order Runge-Kutta scheme. Spatial fluxes are calculated using the total variation diminishing Lax-Friedrichs flux-splitting approach (TVD-LF). In order to minimise wave dissipation due to numerical effects, the grid spacing is chosen such that there are $> 150$ grid points per wavelength of the driven Alfvén waves (denoted by $\lambda_d$). For example, in most simulations, a distance of $100$ ($=50\lambda_d$) has been sampled by $16384$ grid points. 

Before Alfvén waves are excited, we set up a time-invariant plasma background as follows. The magnetic field is aligned with the $x-$axis and has a unit magnitude ($\mathbf{B} = \mathbf{\hat{x}}$). The uniform density and pressure are obtained by setting the Alfvén speed ($v_\mathrm{A}$) and plasma $\beta$ to $1$ and $0.1$, respectively.

\begin{figure*}
    \centering
    \includegraphics[width=0.49\linewidth]{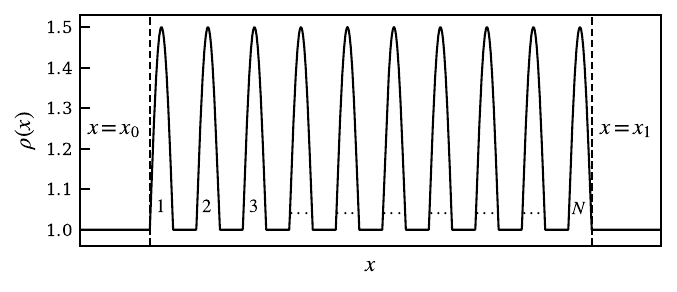}
    \includegraphics[width=0.49\linewidth]{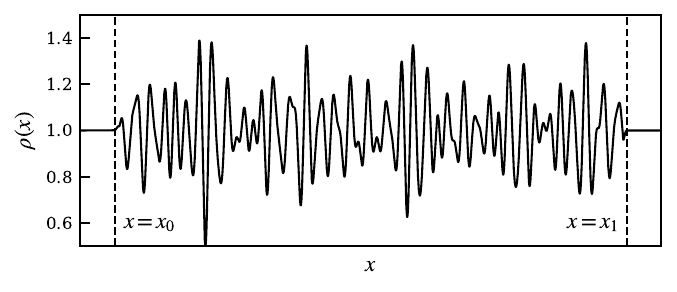}
    \caption{Example density profiles with added inhomogeneities (localised between $x_0 \leq x \leq x_1$) used to study wave reflections and trapping. \textbf{(Left)} Identical blocks of density enhancements with widths $\lambda_s/2$; the gaps between each block is also $\lambda_s/2$, resulting in an overall structuring length scale of $\lambda_s$. \textbf{(Right)} Continuous and randomly varying density enhancements where the mean structuring length scale is $\lambda_s$.}
    \label{fig:density_profiles}
\end{figure*}
Localised density inhomogeneities are then added on top of this background. We consider two cases; (1) a series of identical, half-sinusoidal enhancements (Fig. \ref{fig:density_profiles}, left panel), and (2) continuous, randomly varying fluctuations (Fig. \ref{fig:density_profiles}, right panel). The density enhancement, $\rho^\prime$, in the former is given by
\begin{equation}
    \rho^\prime \left(x\right) =  \sum_{k = 0}^{k = N - 1}\begin{cases} 
      \sin \left(\frac{2\pi x}{\lambda_s} \right) &, k\lambda_s \leq x \leq \left(k + \frac{1}{2}\right)\lambda_s\\
      0 &, \mathrm{otherwise} 
   \end{cases} \label{density_discrete}
\end{equation}
where $N$ is the number of identical density enhancement blocks that repeat over a length scale $\lambda_s$. 

On the other hand, for the randomly varying density inhomogeneities, we define $\rho^\prime$ as
\begin{equation}
\rho^{\prime}\left(x\right) = \sum_{k = 1}^{k = M} a_k \sin\left(\frac{2\pi}{\lambda_k}x + \phi_k\right). \label{density_random}
\end{equation}
Here, we sum over $M$ sinusoids with randomly generated amplitudes, $a_k$, wavelengths, $\lambda_k$, and phases ($\phi_k$). The wavelengths are randomly selected from a Gaussian distribution with a mean value of $\lambda_s$ and the amplitudes and phases are drawn randomly from a uniform distribution. 

In order to ensure that the density profile remains continuous, a smoothing function is used at the start and end of the random enhancements (see lines $x=x_0$ and $x=x_1$ in Fig.~\ref{fig:density_profiles}). We ensure that the density inhomogeneities are localised in space, e.g. between $x_0$ and $x_1$. For $x < x_0$ and $x > x_1$, the background density is uniform, ensuring that we have a clear distinction between forward-propagating incident waves and backward-propagating reflected waves, when a sufficiently short Alfvén wave packet is excited.

The density inhomogeneities are also scaled using a factor, $\alpha$, subsequently referred to as the density contrast. We have
\begin{equation}
    \rho\left(x\right) = \rho_0 + \alpha \rho^{\prime}\left(x\right). \label{total_density}
\end{equation}
This allows us to consider the effects of inhomogeneities with different magnitudes relative to the background density.

Alfvén waves with a period $P_d=2\pi/\omega_d$ are excited at the lower boundary by locally modifying the perpendicular components (aligned with the $y-$ axis in our simulations) of velocity and magnetic field as follows.
\begin{align}
    v_\perp &= A_0 \sin \omega_d t, \label{v_driver}\\
    B_\perp &= -\frac{v_\perp}{\sqrt{\rho}} \label{B_driver}.
\end{align}
Consequently, the injected waves have a wavelength $\lambda_d = 2\pi v_\mathrm{A}/\omega_d$. Here, $A_0$ is the transverse perturbation amplitude, set to $10^{-8}v_\mathrm{A}$, where $v_\mathrm{A}$ is the background Alfvén speed. This very small amplitude ensures that the waves behave linearly in our computational domain. At the boundary, we have implemented outflowing boundary conditions by linearly interpolating all plasma quantities from the computational cells to the ghost cells, thus minimising reflections from the boundary.

The model described above can be further modified to study the effects of a background wind. In that case, we model the lower boundary as the source of a plasma outflow with speed $v_\mathrm{bg}$. This can be either sub-Alfvénic ($M_\mathrm{A} = v_\mathrm{bg}/v_\mathrm{A} < 1$) or super-Alfvénic ($M_\mathrm{A} > 1$). In either case, the Alfvén wave propagation speed is modified to $v_\mathrm{bg} + v_\mathrm{A}$. In the presence of a background flow, the density inhomogeneities are no longer stationary. In this case, the density enhancements are also injected from the lower boundary for the wind to advect them before Alfvén waves are introduced to the system.

\subsection{Methods and analysis}
For all our simulations, we quantify the energy carried by an Alfvén wave using the field-aligned component of the Poynting flux, denoted by $S_n$ (\ref{eq:poynting_flux}), which, in general, is a function of both space and time. We have
\begin{equation}
    S_n\left(x, t\right) = -\frac{1}{\mu_0}\left(\mathbf{B_\perp} \cdot \mathbf{v_\perp} \right)B_\parallel + \frac{1}{\mu_0}\left( \mathbf{B_\perp} \cdot \mathbf{B_\perp}\right)v_\parallel.\label{eq:poynting_flux}
\end{equation}
The second term in (\ref{eq:poynting_flux}) is the flux contribution carried by a background wind, which has zero contribution in the stationary case. The first term, on the other hand, always remains non-zero as the Alfvénic fluctuations are guided by the magnetic field. Noting that the parallel and perpendicular directions correspond to the $x-$ and $y-$ axes respectively in our simulations, (\ref{eq:poynting_flux}) becomes
\begin{equation}
    S_n\left(x, t\right) = \frac{1}{\mu_0}\left(-B_yv_yB_x + B_y^2v_x \right).
\end{equation}
The time integral of the flux at any height $x = x_0$ is the energy passing through a surface described by $x_0$. In order to estimate the amount of reflected and transmitted wave energy, we consider the time-integrated Poynting flux through two surfaces. The first surface is close to the inner boundary of the computational domain which allows us to track the amount of incident and reflected wave energies. On the other hand, the second surface is close to the outer boundary of the domain, which tracks the transmitted energy. If the driver is switched on for $0 \leq t \leq t_1$ and the simulation stops at $t = t_2$, then these energies are given by
\begin{align}
    E_\mathrm{in} &= \int_0^{t_1} S_n \left(x_l, t\right)\, dt, \label{eq:e_in} \\
    E_\mathrm{ref} &= \int_{t_1}^{t_2} S_n\left(x_l, t\right)\, dt, \label{eq:e_ref} \\
    E_\mathrm{out} &= \int_0^{t_2} S_n \left(x_u, t\right)\, dt. \label{eq:e_out}
\end{align}
We can then calculate reflection and transmission coefficients using (\ref{eq:ref_coeff_sim}) and (\ref{eq:trans_coeff_sim}) respectively.
\begin{align}
    R &= \frac{E_\mathrm{ref}}{E_\mathrm{in}}, \label{eq:ref_coeff_sim}\\
    T &= \frac{E_\mathrm{out}}{E_\mathrm{in}}. \label{eq:trans_coeff_sim}
\end{align}
Throughout the paper, wherever possible, we have chosen to present our results in dimensionless quantities, e.g., wave energies in terms of reflection and transmission coefficients, lengths in terms of the Alfvén wave wavelength, speeds in terms of the Alfvén speed, and so on. This not only ensures generality and scalability of the results across different solar and astrophysical conditions, but also facilitates direct comparison with theoretical models and previous studies.

\section{Results} \label{section_results}
\begin{figure*}
    \centering
    \includegraphics[width=0.49\linewidth]{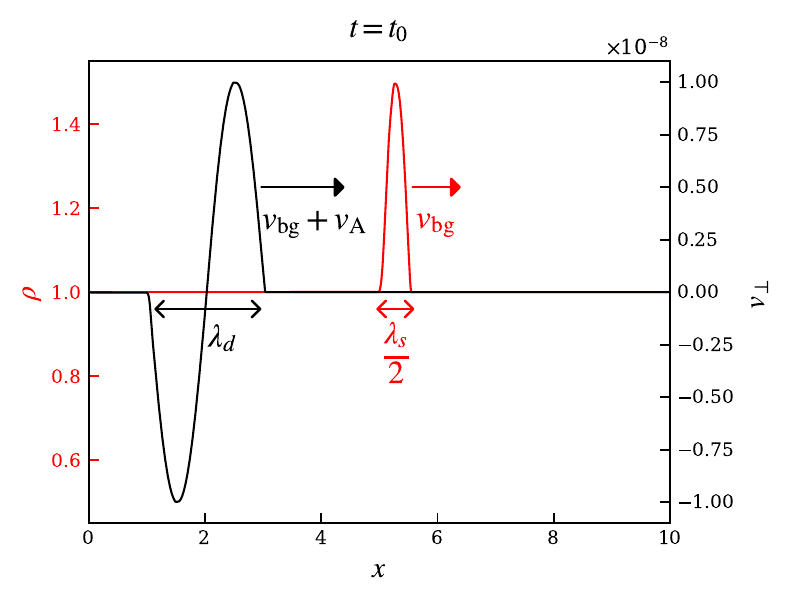}
    \includegraphics[width = 0.49\linewidth]{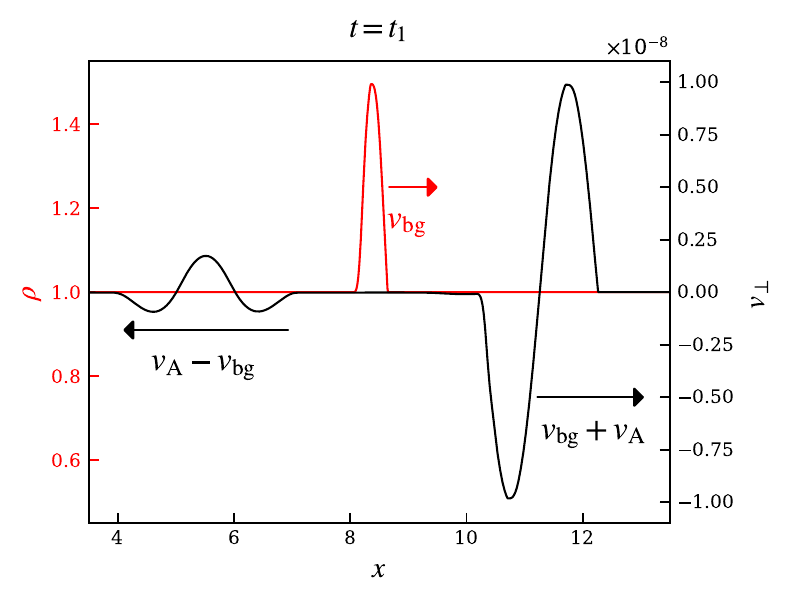}
    \caption{Schematic of the simulation setup -- an Alfv\'en wave packet (shown in black) is injected into the computational domain (left) which interacts with a density inhomogeneity (shown in red) and produces a backward-propagating reflected wave (right). In our simulations, the parameter space study consists of varying $\lambda_s$ and $v_\mathrm{bg}$ to understand their effects.}
    \label{fig:simulation_schematic}
\end{figure*}

Waves spontaneously reflect within a background medium when there are spatial gradients in the wave speed along the direction of propagation. As a result, wave reflection from a varying background density is ubiquitously observed in physical systems. The amount of reflected wave energy is determined by the constructive or destructive interference of incident and reflected waves. As a result, the relative size between the density structuring length scale ($\lambda_s$) and the wavelength of the driven waves ($\lambda_d$) becomes a crucial parameter. It is with this in mind, that in our first set of simulations, we study the effects of $\lambda_s$ on the reflection of a single-wave Alfvén wave packet as it propagates through an isolated block of density enhancement. We perform a parameter study in both a static setup and with a background wind. A schematic of the simulation setup is provided in Fig. \ref{fig:simulation_schematic}. Our goal here is to estimate the optimal conditions for wave reflection, which can be used to study wave trapping in subsequent sections.

\subsection{Wave reflection due to a single density enhancement -- stationary versus non-stationary backgrounds}
\begin{figure}
    \centering
    \includegraphics[width=\linewidth]{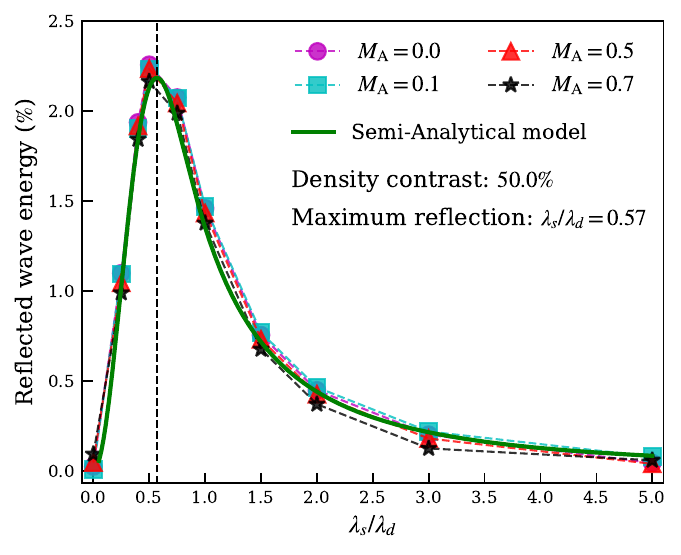}
    \caption{Amount of reflected energy ($R$) as a function of density structuring length scales normalised by the Alfv\'en wave wavelength ($\lambda_s/\lambda_d$). The dashed lines with markers denote the cases with varying background wind speeds. The solid green line denotes the reflection coefficients calculated using a semi-analytical model, showing good agreement with the simulations.}
    \label{fig:single_wave}
\end{figure}

In the first set of simulations that we perform, we not only corroborate the results from \cite{pascoe2022propagating} in a stationary background, but also study the effects of a background plasma flow. With the setup shown in Fig. \ref{fig:simulation_schematic}, we vary the density enhancement length scale ($\lambda_s$), and keep the Alfvén wavelength and density contrast fixed to $\lambda_s$ and $50\%$ of the background, respectively. Additionally, we modify the background wind speed such that the Alfvén Mach numbers are $0.1$, $0.5$ and $0.7$ to understand their effects; $M_\mathrm{A} = 0$ denotes the static case. The results are summarised in Fig. \ref{fig:single_wave}. We have plotted the amount of reflected wave energy against the density enhancement length scale normalised by the Alfvén wave wavelength. The results for each wind speed are shown with different symbols. We also include the results of a semi-analytical model which is discussed in Section \ref{section_semi-analytical_model}.

There are two conclusions that we can derive from Fig. \ref{fig:single_wave}. First, a background wind has minimal effect on the reflection coefficient. Even though a wind contributes positively to the Poynting flux, the factors by which the incident and reflected energies increase are the same, which gives the same reflection coefficient for any given density enhancement length scale. This effect may be due to the fact that the interaction time between an Alfvén wave packet and an enhanced density block remains fixed; the former propagates with a speed $v_\mathrm{bg} + v_\mathrm{A}$, while the latter gets advected with $v_\mathrm{bg}$. The reflected wave packet, on the other hand, must compete against the flow as it travels towards the lower boundary with a speed equal to $v_\mathrm{A} - v_\mathrm{bg}$. As a result, the wind speed controls the direction towards which the reflected wave travels -- if it is sub-Alfvénic, the wave travels towards the lower boundary, while it travels towards the upper boundary when the wind exceeds the Alfvén speed. In any case, a background wind has the effect of slowing down the reflected waves, which increases the travel time back to the source and increases the opportunities for wave energy dissipation.

Second, non-zero reflection coefficients are obtained for all density enhancement lengths, but reflections are more significant for some length scales. The peak in the reflection coefficient spectrum occurs when the density structuring length scale is approximately half the Alfvén wave wavelength, i.e, $\lambda_s \sim \lambda_d/2$, seen across all background wind speeds. Such scale selectivity has previously been reported in \cite{pascoe2022propagating} for Alfvén waves and in \cite{yuan2015evolution} for magnetoacoustic wave pulses. In the following section, we describe a semi-analytical model to explain the scale selectivity.

\begin{figure}
    \centering
    \includegraphics[width=\linewidth]{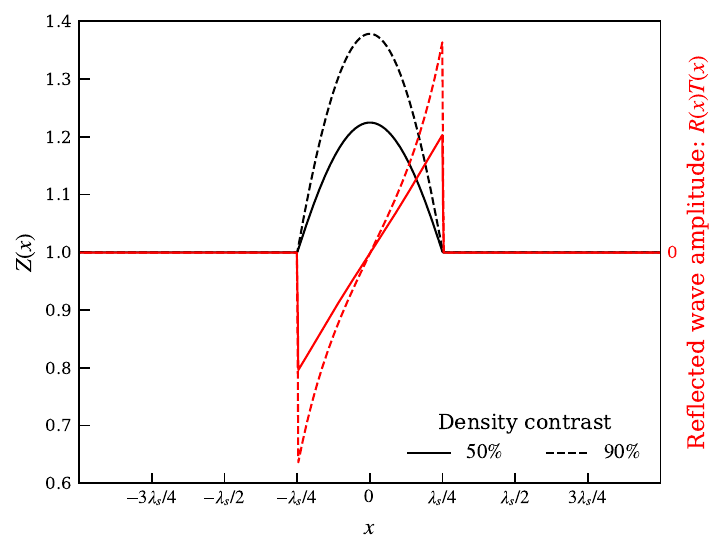}
    \caption{Spatial profile of impedance (black lines) due to density inhomogeneities in the medium. Each point in the domain acts as a source of a reflected wave, the amplitude of which is scaled by the local reflection and transmission coefficients, shown as red lines in the figure. A negative reflected wave amplitude corresponds to a phase shift by $\pi$ with respect to the incident wave. The dashed lines denote cases for a larger density contrast.}
    \label{fig:analytical_impedance}
\end{figure}

\begin{figure*}
    \centering
    \includegraphics[width=\linewidth]{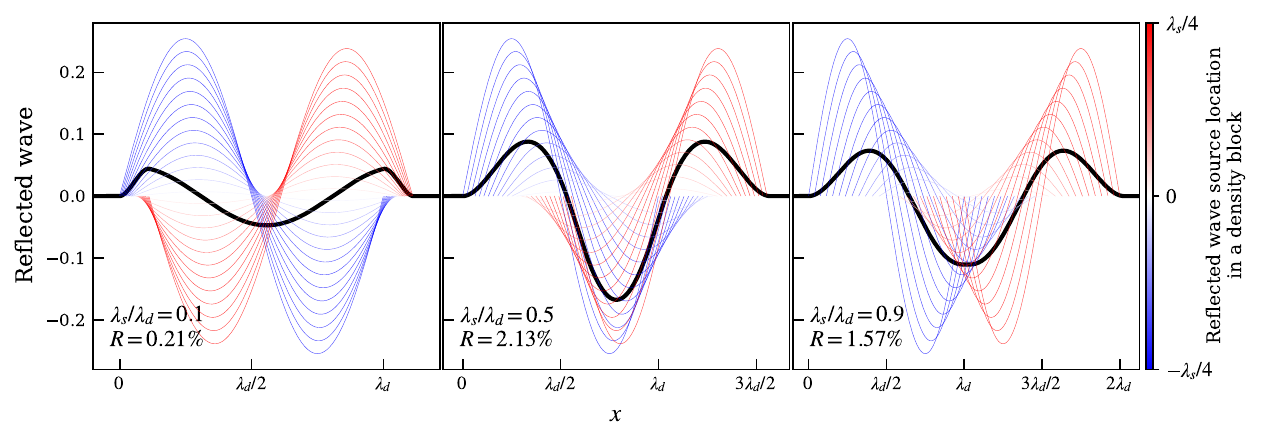}
    \caption{Reflected waves (shown in black) for different density structuring length scales. The blue and red lines, drawn on a different scale, are the constituent reflected waves, coloured according to their site of reflection -- blue denotes closer to the leading edge while red denotes the trailing edge in the density enhancement block of width $\lambda_s/2$; the colour bar corresponds to the $x-$axis in Fig. \ref{fig:analytical_impedance}.}
    \label{fig:reflected_waves_analytical}
\end{figure*}

\subsubsection{A Semi-Analytical Model to Understand Scale Selectivity} \label{section_semi-analytical_model}
We note that the Alfvén waves excited in our simulations are incompressible owing to their small perturbation amplitude. This means we can assume that the background does not change as the waves propagate through it. With this assumption, we construct a reduced-physics model based on the interference of waves to understand why some density inhomogeneity length scales are preferred over others for efficient reflections. 

We consider each point in the background as a potential source of reflected waves which will propagate in a direction opposite to the incident wave. The reflected waves can then interfere constructively or destructively depending on their relative phase shift. The key parameters to consider here are the reflection and transmission coefficients, denoted by $R$ and $T$ respectively. We have
\begin{align}
    R &= \frac{Z\left(x_1\right) - Z\left(x_2\right)}{Z\left(x_1\right) + Z\left(x_2\right)}, \label{reflection_coefficient}\\
    T &= \frac{2Z\left(x_1\right)}{Z\left(x_1\right) + Z\left(x_2\right)} \label{transmission_coefficient}.
\end{align}
The above equations are obtained when a wave propagates between media $1$ and $2$, denoted by points $x_1$ and $x_2$, respectively. Wave reflections occur only when there is a change in impedance between the two points. In other words, $Z\left(x_1\right) \neq Z\left(x_2\right)$, where $Z(x)$ is defined as
\begin{equation}
    Z\left(x\right) = \rho \left(x\right) v\left(x\right) = \rho\left(x\right) \left(v_\mathrm{bg} \left(x\right) + v_\mathrm{A} \left(x\right) \right). \label{eq:impedance}
\end{equation}
To model wave reflections due to the entire density enhancement, we discretise the enhancement. We then label each neighbouring point in the domain as $x_1$ and $x_2$, such that $\left|x_1 - x_2 \right| \ll 1$ and calculate the local reflection and transmission coefficients. As an incident wave reaches that point, the reflected wave is obtained by scaling the incident wave amplitude by these coefficients. In Fig. \ref{fig:analytical_impedance}, we show the impedance and the reflected wave amplitude -- a consequence of the local reflection and transmission coefficients -- for an enhanced density block in the medium. As the density contrast is increased, the impedance gradient also increases, which increases the reflected wave amplitude. Furthermore, we see that the maximum contribution comes from the edges of the density enhancement, where the gradients in impedance are the steepest. When the above procedure is repeated for all points in the domain, we obtain a family of reflected waves, which collectively form the combined reflected wave. This is depicted by the black line in Fig. \ref{fig:reflected_waves_analytical}. Additionally, the individual components are shown in coloured lines, drawn on a different scale for clarity. In our model, we have also accounted for the phase difference between each individual reflected wave, which is picked up due to different travel times from their respective reflection points. 

In order to calculate the energies, we again use the Poynting flux (\ref{eq:poynting_flux}). In our calculation, we use the Alfvén relation ($v_\perp = -B_\perp/ \sqrt{\rho}$), which yields $v_\perp = -B_\perp$ because the background density is set to unity. Consequently, the energy reduces to a time integral of the squared wave component, denoted by $v_\perp$ in (\ref{semi_analytical_energy}).
\begin{equation}
    E = \int v_\perp^2\, dt \label{semi_analytical_energy}
\end{equation}

In order to explain the density length scale selectivity observed in our simulations, we initialise our semi-analytical model with the same configuration as in Fig. \ref{fig:simulation_schematic}, but without a background wind. This computationally inexpensive semi-analytical model enables us to calculate reflection coefficients for an arbitrary resolution in $\lambda_s$. The spectrum is shown by the solid green line in Fig. \ref{fig:single_wave}. We note that the reduced-physics model is in agreement with our 1.5-D MHD simulations. We therefore conclude that the dominant effect causing some length scales to reflect more can be understood in terms of the interference of reflected waves. For a stand-alone Alfvén wave pulse interacting with a $50\%$ density enhancement, the maximum energy reflected is $\sim 2.2\%$ when $\lambda_s/\lambda_d = 0.57$. We note that this is sensitive to the spatial form of the density enhancement.

As mentioned above, when we examine our density enhancements, we find that the highest contribution to the reflected wave comes from the edges, where gradients are largest. Depending on the separation of the edges (the density enhancement length scale, $\lambda_s$), reflected waves from these sites may be out of phase. In order to maximise reflection, we require the reflected wave troughs to overlap. This will  maximise the amount of constructive interference. From our simulations, we find that this only happens when $\lambda_s \sim \lambda_d/2$ (shown in the middle panel of Fig. \ref{fig:reflected_waves_analytical}). For other values of $\lambda_s$, there is either a net destructive interference -- between a crest and a trough when $\lambda_s$ is small (left panel in Fig. \ref{fig:reflected_waves_analytical}), or an ineffective constructive interference -- between successive troughs when $\lambda_s$ is made larger (right panel in Fig. \ref{fig:reflected_waves_analytical}).

\begin{figure}
    \centering
    \includegraphics[width=\linewidth]{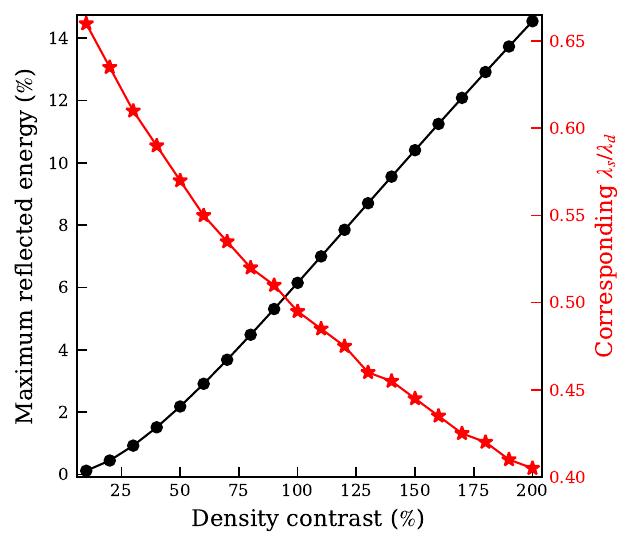}
    \caption{The effect of density contrast on the amount of reflected energy (shown by black circles). The red stars denote the conditions in parameter space for which reflected energy is maximised for a given density contrast.}
    \label{fig:analytical_density_contrasts}
\end{figure}

Having shown that the amount of reflected energy depends on the density enhancement length scale, we now investigate the role of the density contrast. Using the semi-analytical model we now vary both the density contrast and the enhancement width to understand their combined effects. We find that the spectra are similar to the previously obtained results, but maximum reflected energy and the optimal conditions have changed; these quantities are plotted in Fig. \ref{fig:analytical_density_contrasts}. As the density contrast is increased, the impedance gradient steepens, resulting in an increase in the reflection coefficient. Furthermore, the propagation speed ($v_\mathrm{A}$) decreases, meaning that waves originating from two neighbouring points will have a larger path difference for a higher density contrast. As a result, optimal constructive interference of the reflected wave packets is obtained for different density enhancement widths. In particular, higher density contrasts reduce the value of $\lambda_s/\lambda_d$ for which the reflected energy percentage is maximal.

\begin{figure*}
    \centering
    \includegraphics[width=\linewidth]{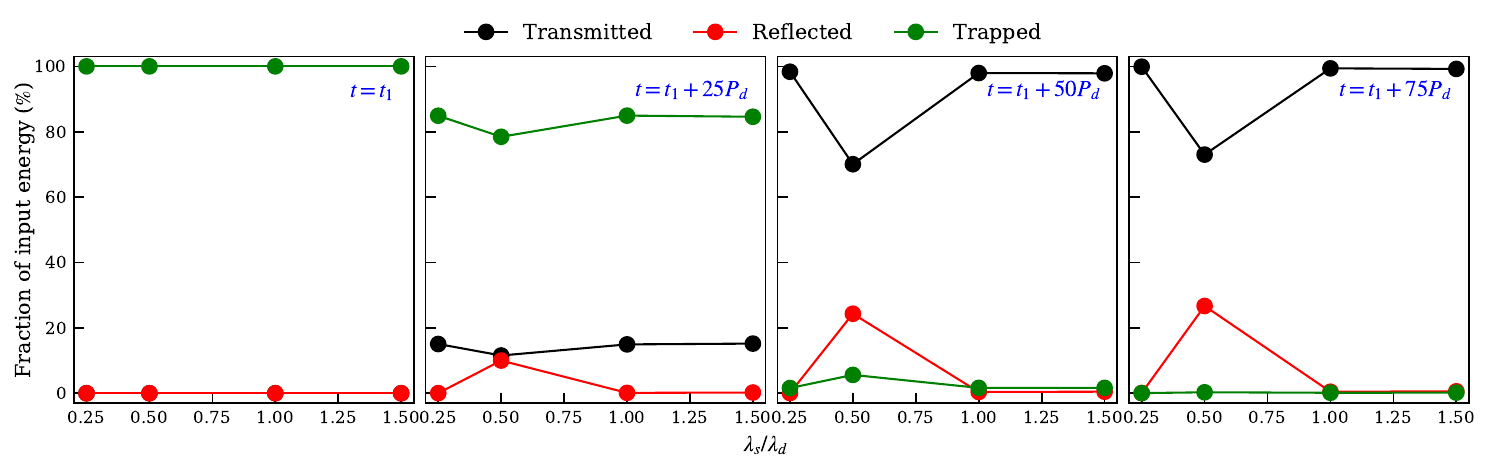}
    \caption{Amounts of transmitted (black), reflected (red) and trapped (green) energy fractions at different times of the simulation for varying random density enhancement length scales ($\lambda_d$); the density contrast is fixed at $20\%$ of the background. The time is measured in terms of wave periods after the driving phase ($t > t_1$).}
    \label{fig:random_length}
\end{figure*}

\begin{figure*}
    \centering
    \includegraphics[width=\linewidth]{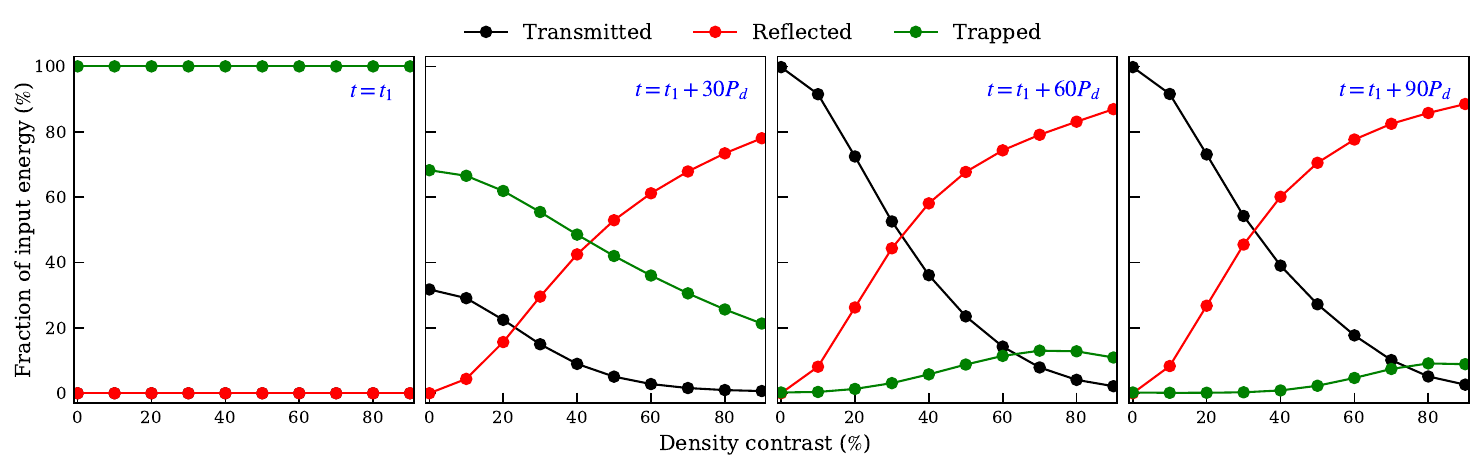}
    \caption{Amounts of transmitted (black), reflected (red) and trapped (green) energy fractions at different times of the simulations for varying density contrasts, while the enhancement length scale is fixed to $\lambda_d/2$ for optimum reflections. The time is measured in terms of wave periods after the driving phase ($t > t_1$).}
    \label{fig:random_density_contrast}
\end{figure*}

\begin{figure}
    \centering
    \includegraphics[width=\linewidth]{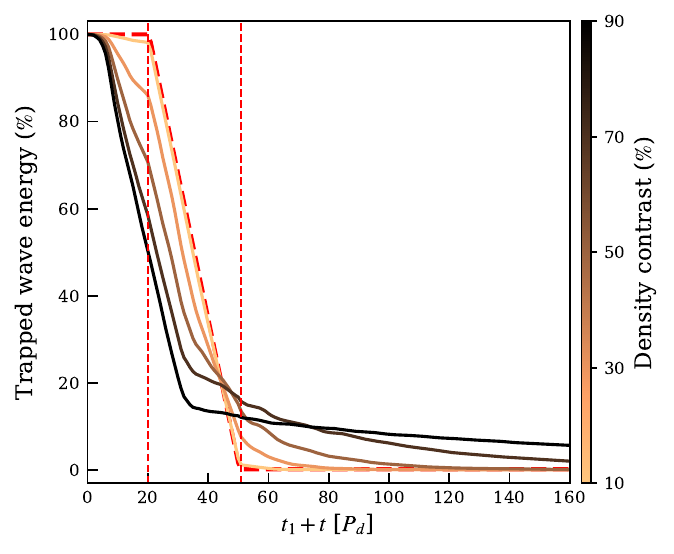}
    \caption{Trapped energy fraction time series for varying density contrasts with optimised density enhancement length scale. Again, the time is measure in terms of wave periods after the driving phase ($t > t_1$). The vertical red dashed lines denote the times at which the leading and trailing wavefronts leave the domain from the upper boundary when there are no density enhancements.}
    \label{fig:random_density_trapping}
\end{figure}

\subsection{Wave reflection and trapping due to random density perturbations}
In this section, we consider a more general case of an Alfvén wave train interacting with a region of randomly generated density enhancements. As with our previous simulations, we start by investigating the effects of the structuring length scale ($\lambda_s$), which is the mean length scale about which the sinusoids are randomly selected in (\ref{density_random}). We vary $\lambda_s$ while keeping the density contrast fixed to $20\%$ with respect to the background. An Alfvén wave train that is $30\lambda_d$ long is injected from the lower boundary. The results are shown in Fig. \ref{fig:random_length}, where we have plotted the transmitted (black), reflected (red) and trapped (green) energies as fractions of the incident energy at different times during the simulations; the trapped energy is calculated by subtracting the reflected and transmitted energies from the input energy and is merely a measure of the wave energy remaining in the domain at a given time.

As soon as the wave driving ends, all the wave energy is within the computational domain. After some more time has elapsed, some energy will be transported through either boundary of the domain. If the associated waves leave through the lower boundary, they contribute to the reflected energy, whereas if they leave through the upper boundary, they contribute to the transmitted energy. In Fig. \ref{fig:random_length}, we once again see a peak in the reflected energy when $\lambda_s \sim \lambda_d/2$. For this case, there is also a significant amount of wave energy trapped in the domain. This is a result of repeated reflections between the density inhomogeneities. Eventually, all the trapped energy leaves the domain from either boundary, since density enhancements are only partially reflective. 

We also perform a further set of simulations in which the density contrast varies and the density structuring length scale is fixed to $\lambda_d/2$. In Fig. \ref{fig:random_density_contrast}, we plot the reflected, transmitted and trapped wave energies for different density contrasts. We immediately see that higher contrasts lead to greater contributions to the reflected energy. Additionally, a higher fraction of wave energy is trapped between the density enhancements, as a greater proportion of the wave energy is associated with repeated reflections. 

In Fig. \ref{fig:random_density_trapping}, we present a more detailed analysis of the wave trapping duration. The red dashed line represents a case with no density enhancements and therefore no (repeated) reflection. Before $t=t_1 + 20 P_d$ (i.e., $20$ wave periods after the driving phase), the first wave front has not yet reached the upper boundary and therefore all of the wave energy remains in the domain. After $t=t_1 + 52P_d$, the final wave front has left the domain (in the uniform density case) and thus all wave energy has been transmitted. On the other hand, for the simulations with density variation, the trapped energy starts to decrease even before the incident wave reaches the upper boundary. This is because some reflected waves are transmitted through the lower boundary first. Beyond $t=t_1 + 20P_d$, some wave energy is also transmitted through the upper boundary and so a sharp decline in the trapped energy content is observed. However, the possibility of repeated wave reflections results in a delay for the waves to completely leave the computational domain. We see that this effect is more pronounced for larger density contrasts.

\subsection{The effects of interaction width between density enhancements and Alfvén waves on their reflection and trapping}

\begin{figure*}
    \centering
    \includegraphics[height = 5.7cm]{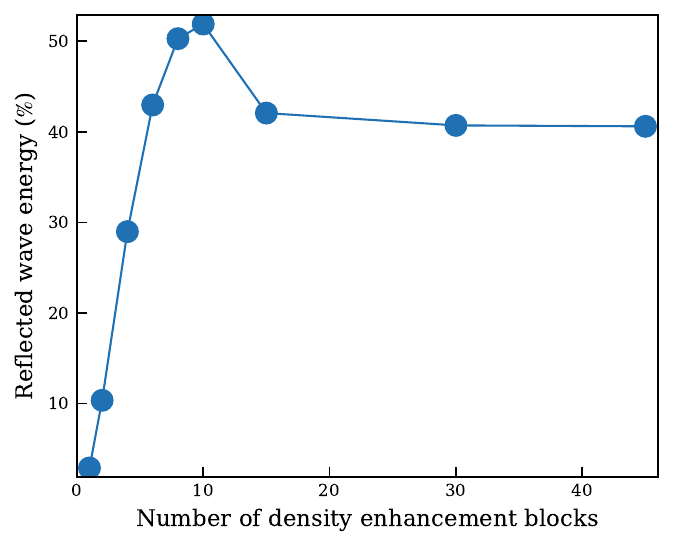}
    \includegraphics[height = 6cm]{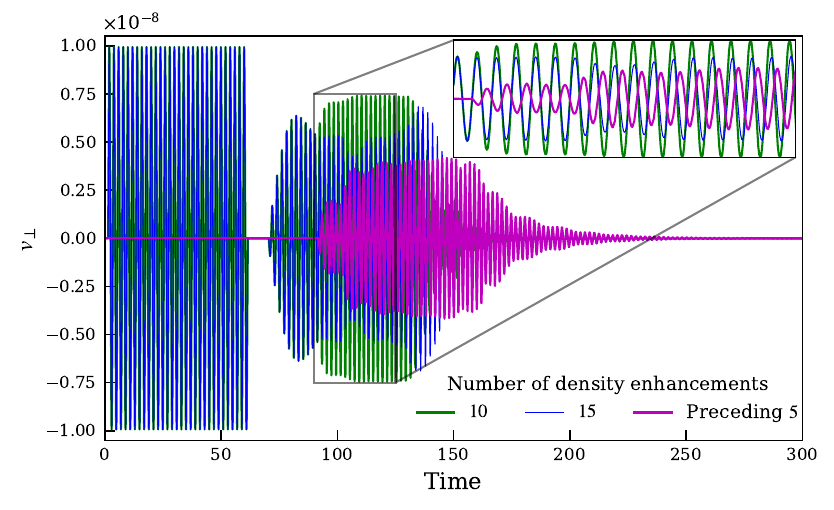}
    \caption{\textbf{(Left)} Reflected wave energy fractions for varying number of density enhancement blocks ($N$). \textbf{(Right)} Time series of the perpendicular velocity component at the lower boundary. The green and blue lines depict cases when there are $10$ and $15$ density enhancement blocks respectively. The magenta line depicts the waves generated due to the additional $5$ blocks (drawn by taking a difference between the blue and green lines), which is out of phase with respect to the waves generated by $10$ density enhancements.}
    \label{fig:ref_vs_blocks}
\end{figure*}

\begin{figure}
    \centering
    \includegraphics[width=\linewidth]{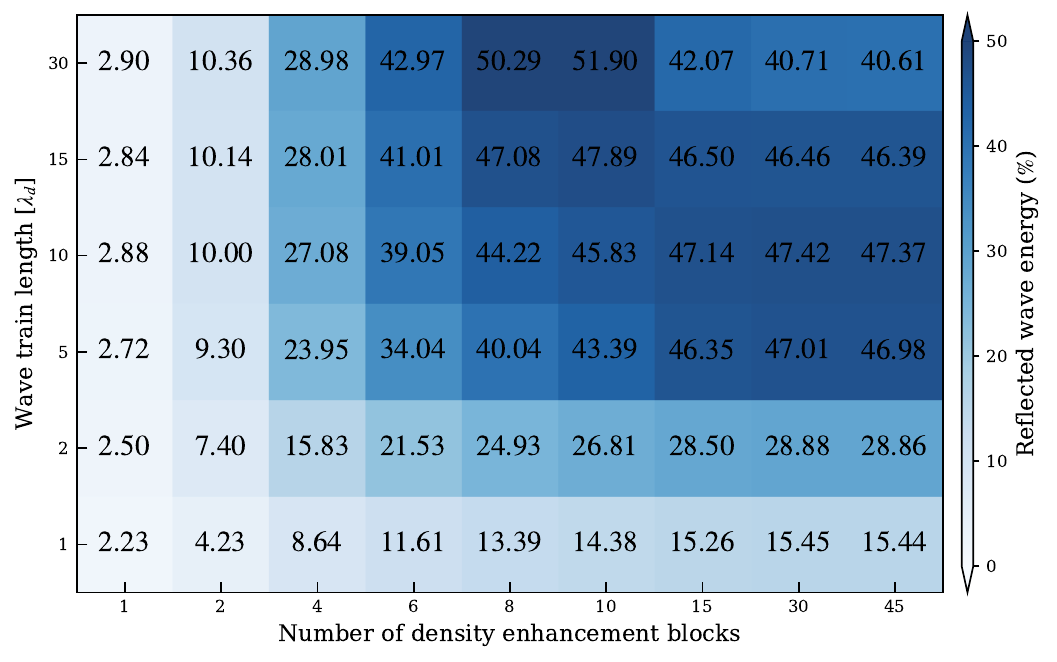}
    \caption{Fraction of reflected wave energy for different simulations with varying input wave train lengths (measured in multiples of $\lambda_d$) and varying numbers of identical density blocks having a mean length scale $\lambda_s$, which is fixed to $\lambda_d/2$.}
    \label{fig:multi-blob-multi-wave}
\end{figure}

\begin{figure}
    \centering
    \includegraphics[width=\linewidth]{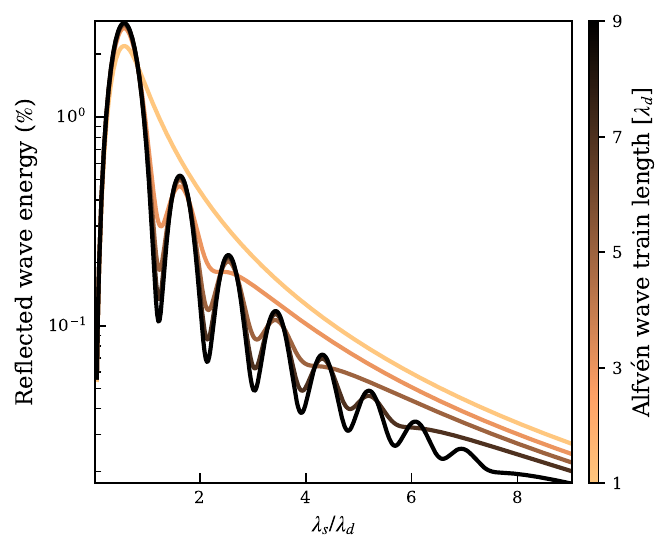}
    \caption{Reflected wave energy spectrum for varying input Alfvén wave train lengths (measured in multiples of $\lambda_d$) interacting with a single density enhancement block with a variable width ($\lambda_s$), using the semi-analytical model.}
    \label{fig:multi_wave_spectra}
\end{figure}

In the previous sections, we have looked at how Alfvén wave reflections are affected by either the density contrast or the density enhancement length scale. Increasing the former usually increases the amount of reflections while the reflection coefficient attains a peak for intermediate values of the density enhancement length scale. Furthermore, we would expect the reflected energy fraction to also depend on the interaction length between the waves and the density enhancement region, which can be done in two ways -- vary the input wave train length, and/or vary the number of density enhancement blocks (the density enhancement profile is similar to the one shown in the left panel of Fig. \ref{fig:density_profiles}). Here, we perform these simulations keeping the density contrast fixed to $50\%$ of the background and the mean structuring length scale fixed to $\lambda_d/2$. While a single density enhancement has only one reflection site, having a series of density enhancements opens new sites for wave reflection. This also promotes wave trapping as reflected waves can undergo further reflections to generate counter-propagating waves in between density enhancements.

As a first set of simulations, we excite an Alfvén wave train that is $30$ wavelengths long, which interacts with some density enhancements having a variable number of blocks (denoted by $N$ in the left panel of Fig. \ref{fig:density_profiles}). The resulting reflected energy fractions are plotted as a function of the number of density enhancement blocks in the left panel of Fig. \ref{fig:ref_vs_blocks}. The energy profile has three phases -- one, a sharp increase when there are fewer density blocks; two, attaining a maximum reflection coefficient for $N \sim 10$; and three, a saturation for large number of density enhancements.

In a scenario such as this, additional density enhancements open new sites of reflection, which increases the reflection coefficient. However, we note that each density enhancement transmits only a fraction of the incident energy, which lowers the amount of energy available for reflection for the following enhancement. As a result, contributions from subsequent density enhancements diminish and we would soon expect a saturation in reflection coefficients. 

Additionally, when the density enhancement length scales become comparable with the incident wave wavelength, interference effects become significant, which is why a peak is attained in the reflection coefficient profile in the left panel of Fig. \ref{fig:ref_vs_blocks}. Waves reflected from subsequent density enhancements are phase shifted with respect to the reflected waves generated from preceding enhancements. For lower numbers of density enhancements, the phase difference is not significant and the waves interfere constructively; for higher number of density enhancements ($N \gtrapprox 10$), the phase difference becomes greater than $\pi/2$ and waves start to interfere destructively. This is shown in the right panel of Fig. \ref{fig:ref_vs_blocks}, where we have compared the two anomalous cases -- $10$ and $15$ density enhancement blocks; the additional reflected waves from the preceding $5$ blocks is shown in magenta, which is largely out of phase with respect to the green line corresponding to $10$ blocks. This results in a drop in the reflection coefficient when the number of density enhancements is further increased from $10$. Therefore, the saturation in Fig. \ref{fig:ref_vs_blocks} can be explained by considering two competing effects -- wave interference, and diminishing contribution due to subsequent density enhancements. 

In Fig. \ref{fig:multi-blob-multi-wave}, we tabulate the reflection coefficients for varying numbers of density enhancements and incident Alfvén wave train length (measured in terms of its wavelength, $\lambda_d$). It is important to note that the dip in the reflection coefficient occurs only for longer wave trains, which is because wave interference becomes unimportant for shorter trains, and only saturation effects are significant. 

As the input wave train length is increased while keeping the number of density enhancement blocks fixed, we observe that the reflection coefficient increases in general, but a saturation is achieved for longer wave trains. Again, the increase becomes more prominent for higher density enhancement blocks. In order to help explain this effect, we invoke the semi-analytical model described in Section \ref{section_semi-analytical_model}, where we consider an Alfvén wave train that is several wavelengths long, instead of a single pulse, interacting with a single density enhancement block.

The length of the reflected wave is determined by the length of the incident wave train, as well as the spatial extent of the density enhancement. This provides increased opportunity for interference between the incident and reflected waves and thus drives more complex behaviour, resulting in a reflection coefficient spectrum as shown in Fig. \ref{fig:multi_wave_spectra}. We show the effects of varying the length of the incident wave train. The peaks and valleys in the spectra, occurring at integer multiples of the first peak or valley, correspond to regions of constructive and destructive interference, respectively. Although the peak energy values do not differ significantly across different incident wave trains, successive peaks for the reflected energies decrease logarithmically for a wave train of a given length.

\section{Discussion and conclusions}\label{section_discussion}
In this paper, we study Alfvén wave propagation in a medium with longitudinal density enhancements to determine the optimal conditions for wave energy trapping. Such density enhancements partially reflect the incident waves, which are consequently trapped in between these enhancements. Although we do not study Alfvén wave energy dissipation, we demonstrate prolonged wave activity in between density enhancements, for a low-$\beta$ plasma such as the solar corona. Using a parameter-space study, we confirm the result of \cite{pascoe2022propagating} that the reflection coefficient maximises when the density enhancement length scale is approximately half of the Alfvén wave wavelength ($\lambda_s = \lambda_d/2$). This scale-selectivity is a net result of wave interference, as each point in the continuum acts as a source of a reflected wave, with each wave having an associated phase shift. We note that wave interference is an important effect when it comes to optimising wave reflection in an array of density enhancements. 

Building on the work done by \cite{pascoe2022propagating} and \cite{yuan2015evolution}, we performed additional studies on the effects of a sub-Alfvénic wind. Since the backward propagating reflected waves travel with a speed equal to $v_\mathrm{A} - v_\mathrm{bg}$, only wave reflections which happen in the sub-Alfvénic regime of the solar wind lead to energy trapping at lower heights. We find that the reflection coefficients remain unaffected by the background wind speed, as the relative velocity, and therefore the interaction time, between the density enhancement and an Alfvén wave pulse remain the same ($=v_\mathrm{A}$). 

Alfvén wave reflection and trapping becomes more complicated in an expanding, gravitationally-stratified solar wind. This is because reflections occur at every point in space -- due to a continuous variation in the density and wave propagation speed -- and also because Alfvén waves are likely to have a broadband spectrum. As the wave propagation speed varies significantly in the solar wind, so does the wavelength of propagating Alfvén waves. At low altitudes, the propagation speed is dominated by the background Alfvén speed. However at higher altitudes, the acceleration of the wind means the background plasma flow is a significant component of the overall propagation speed. For typical solar wind conditions, and for waves with a period of $5$ minutes, we find wavelengths range between $0.05\,R_\odot$ and $0.2\,R_\odot$ (for heliocentric distances $< 10\, R_\odot$). This means that the optimal conditions for wave reflections ($\lambda_s \approx \lambda_d/2$) are modified throughout the wind. An additional, important effect arises due to the acceleration of the solar wind. The non-uniform wind speed means that the interaction time between an Alfvén wave packet and a density enhancement block depends on their location in space. As such, we expect a departure from the $\lambda_s \approx \lambda_d/2$ condition where the wind is accelerating.

In a system with more than one density enhancement, additional enhancements provide additional reflection sites. This results in a net increase in the reflection coefficient as the number of density enhancements is increased. This is the primary effect in our simulations. However, once a threshold of the number of density enhancements is reached (this threshold depends on the specifics of the density enhancements, such as their amplitudes and widths), secondary effects due to wave attenuation and wave interference become significant. As the density enhancements are partial reflectors, the amount of energy available for reflection from the subsequent density enhancement decreases, resulting in a saturation of the reflection coefficient profile. This behaviour is further modified by the fact that reflected waves from additional density enhancements are phase shifted with respect to the ones generated from preceding density enhancements. This has the effect of lowering the net reflection coefficient when the reflected waves are largely out of phase. Eventually, we find saturation of the reflected wave energy governed by the effects discussed above.

The semi-analytical model discussed in Section \ref{section_semi-analytical_model} demonstrates that wave interference is a key process in determining the reflected wave energy. This causes the scale selectivity -- why maximum reflection is observed for specific density inhomogeneity length scales. This may become a useful tool to understand wave energy retention for more complex scenarios. Although each point in a density enhancement acts as a source of reflected waves (with amplitudes dependent on the gradients in local impedance), they must interfere constructively to increase the reflected wave energy. The semi-analytical model described in the paper, because of its simplicity, avoids running long and expensive MHD simulations and is able to predict reflection coefficients accurately. We note that this assumes that waves do not affect the background medium. This is true for linear Alfvén waves. Furthermore, the semi-analytical model does not take into account the effects of wave trapping, as each reflection site reflects only once to produce the family of reflected waves.

In our paper, we have demonstrated that wave activity can be prolonged when there are field-aligned density enhancements in the solar wind. However, the amount of trapped energy largely depends on the relative sizes of the density inhomogeneity length scales and the Alfvén wave wavelength. We have shown that this can be attributed to effects of wave interference. Through a parameter-space study, we have also provided conditions for maximal wave reflection in an extended region of density enhancements to identify an upper limit to the reflection coefficient. This is caused by reflected waves interfering destructively for a sufficiently high number of density enhancements. 

Finally, we note that the simulations and semi-analytical model presented here only consider a simple one-dimensional setup. In the solar wind, additional effects such as field expansion, spherical geometry, gravitational stratification, solar wind acceleration, intermittent heating events and cross-field variations couple to create a much more complex system. However, our simple analysis provides a framework which could be used to describe wave dynamics in more complicated settings, without necessarily relying on large-scale numerical simulations.

\section*{Acknowledgements}
AK received funding from the Leverhulme Trust (RPG-2021-114). TAH received funding from the Carnegie Trust grant number RIG013386. PP acknowledges support by the Italian PRIN 2022, project 2022294WNB entitled ``Heliospheric shocks and space weather: from multispacecraft observations to numerical modeling”, funded by Next Generation EU, within Piano Nazionale di Ripresa e Resilienza (PNRR), Missione 4 ``Istruzione e Ricerca” - Componente C2 Investimento 1.1, `Fondo per il Programma Nazionale di Ricerca e Progetti di Rilevante Interesse Nazionale (PRIN)'. IDM received funding from the Research Council of Norway through its Centres of Excellence scheme, project number 262622.

\section*{Data availability}
The source code for the numerical simulations and analysis routines implemented during the preparation of this article will be available from the \href{https://research-portal.st-andrews.ac.uk/en/publications/alfv%C3%A9n-wave-propagation-reflection-and-trapping-in-the-solar-wind}{University of St Andrews research repository}.

\bibliographystyle{mnras}
\bibliography{bibliography}

\bsp	
\label{lastpage}
\end{document}